\documentstyle[epsf]{aipproc}  
\def\psim{\lower.5ex\hbox{$\; \buildrel \propto \over\sim \;$}}  
\def\gtrsim{\lower.5ex\hbox{$\; \buildrel > \over\sim \;$}}  
\def\ggrsim{\lower.5ex\hbox{$\; \buildrel \gg \over\sim \;$}}  
\def\lesssim{\lower.5ex\hbox{$\; \buildrel < \over\sim \;$}}  
 
\def\g2{\gamma_2}

\def\e{{\epsilon}}  
  
\def\E{{\cal E}} 
\def\m{{\cal M}} 
  
\begin{document}  
\title{Gamma Ray Bursts, Cosmic Ray Origin,\\ 
\vskip0.05in and the Unidentified EGRET Sources}  
\author{Charles D. Dermer\thanks{Work supported by the Office of Naval Research. Presentation at Second Rome Workshop on Gamma Ray Bursts in the Afterglow Era (17-20 October 2000). Shortened version to appear in the Proceedings of the Heidelberg 2000 High-Energy Gamma-Ray Workshop, ed. F. A. Aharonian and H. V\"olk  (AIP: New York).}} 
\address{Naval Research Laboratory, Code 7653, Washington, DC  20375-5352 USA} 
 
%\lefthead{LEFT head}  
%\righthead{RIGHT head}  
\maketitle  
  
\begin{abstract}  
Statistical arguments show that the volume- and time-averaged kinetic power of GRBs and fireball transients (FTs) into an L$^*$ galaxy like the Milky Way is at the level of $10^{40}$ ergs s$^{-1}$. This number, though with wide uncertainties related to the internal or external shock efficiency, is sufficient to power hadronic cosmic rays observed locally. 
 
The release of energy by the high-mass progenitor stars of GRBs and FTs is sufficient to power the ultrahigh energy cosmic rays, as already shown by Waxman and Vietri in 1995.  It is sufficient to power the cosmic rays above the knee of the cosmic-ray spectrum. Indeed, all hadronic cosmic rays could originate from the high-mass ($\gtrsim 100$ M$_\odot$) stars that collapse to black holes, in the process forming GRBs and FTs. This source class represents a new solution to the problem of cosmic-ray origin. 
 
The $\sim 10^4$-$10^7$ black holes made by these stars could make their presence known by radiating as they accrete from the ISM, by microlensing background radiations, and by forming luminous binary systems. Some unidentified EGRET sources could be isolated black holes that accrete from the ISM. Better imaging and sensitivity with GLAST and TeV observatories will test this model for the unidentified $\gamma$-ray sources, and this theory for cosmic-ray origin. 
\end{abstract}  
  
\section{Introduction}  
 
Fireball transients are those explosive events that propel a significant fraction of their ejecta kinetic energy in the form of relativistic baryonic outflows. FTs include dirty ($\Gamma_0 \lesssim 300$), loaded ($\Gamma_0\sim 300$), and clean fireball ($\Gamma_0 \gtrsim 300$) subclasses \cite{dcb99}, of which BATSE is most sensitive to those explosions with initial bulk Lorentz factors $\Gamma_0 \sim 300$. The dirty and clean fireball classes of exploding stars are difficult to discover because of design and sensitivity limitations of detectors flown to date\cite{dcb99,dbc99}, though the X-ray flashes detected with Beppo-SAX \cite{hei00} could be representatives of the dirty fireball class. HETE-II, Swift, and GLAST can all be expected to make progress on this front, though a lobster-eye X-ray telescope seems most promising for discovering dirty fireballs. 
 
In a recent paper \cite{der00}, I deduced the time- and volume-averaged kinetic power of GRBs and fireball transients (FTs) in the Milky Way on the basis of a statistical treatment employing the external shock model of GRBs \cite{rm92,mr93} for both the prompt $\gamma$-ray luminous and afterglow phase, performed by B\"ottcher and myself \cite{bd00}. There I found that the FT power into the galaxy was within an order of magnitude of the inferred average Galactic cosmic-ray power, and I considered the possibility that the high-mass stellar progenitors of FTs were the sources of the cosmic rays (CRs) above and below the knee of the CR spectrum. I also discussed observational difficulties in the scenario advanced by Ginzburg and Syrovatskii \cite{gs64} that hadronic CRs are powered when supernovae collapse to neutron stars. 
  
Although $\gamma$-ray astronomy was predicted to solve the CR origin problem (see \cite{wee99} for a recent assessment), evidence for CR source origin still rests primarily upon arguments about available sources of power, which in turn hinge upon statistical analyses. The numerical simulation of GRB statistics \cite{bd00} that was used to derive the FT power in the Milky Way modeled detector response and generic GRB spectral and temporal behavior to fit jointly large samples of GRB properties measured with BATSE, namely the peak photon energies $\e_p^0$ of the prompt $\nu F_\nu$ spectra, the peak count rates, and the $\gtrsim 1$ s $t_{50}$ durations. The predicted redshift distribution of GRBs can be used to test the model, but will require several score GRB redshifts from HETE-II and/or Swift. The redshift sample should preferably come from a single mission in order to minimize triggering differences and calibration uncertainties between detectors.   
 
Rather than review the numerical results for a third time (see \cite{dc99,der00a}), it seems more fruitful to devote this space to an analytic statistical model that confirms in all respects the numerical results.  We gain clarity at the expense of the accuracy found in \cite{bd00}. 
 
In section II, the standard argument that only supernova explosions are frequent and energetic enough to power the galactic cosmic rays is recited, while adumbrating certain weaknesses in the underlying assumptions.  The statistics of cosmological GRBs are treated in Section III. Depending on shock efficiency during the prompt phase, the power of the high-mass progenitors into an L$^*$ galaxy is evaluated. GRBs and FTs are shown to be capable of injecting sufficient energy into the ISM to power the hadronic CRs, as discussed in Section IV. Potentially observable consequences of the $10^4$-$10^7$ black holes that are formed from FTs in the Galaxy are considered in Section V, including a possible solution to the origin of some of the unidentified EGRET sources. The lineage of stars that explode as different novae types is sketched in Section VI. Section VII provides a summary and a brief discussion of the acceleration and adiabatic loss problem, that will be treated in more detail elsewhere \cite{dh00}. 
 
\section{Cosmic Ray and Supernova Power} 
 
Observations show that TeV electrons are accelerated by SN remnants (SNRs), but there is as yet no direct observational evidence for hadronic CR acceleration by SNRs \cite{der00,dp99}. Perhaps the most compelling argument that hadronic CRs are powered by SNRs is the claim that only SNe inject sufficient power into the Galaxy to provide the observed CR energy density \cite{gs64,gai90}. The local energy density of CRs is $u_{CR} \sim 1$ eV cm$^{-3} \approx 10^{-12}$ ergs cm$^{-3}$. The required CR power is thus $L_{CR} \approx u_{CR} V_{gal}/t_{esc}$, where $V_{gal}$ is the effective volume of the Galaxy from which CRs escape on a timescale $t_{esc}$. If CRs are trapped in a disk of 15 kpc radius and 100 pc scale height, then $V_{gal} \approx 4\times 10^{66}$ cm$^{3}$.  Observations of the light elements Li, Be, and B that are formed through spallation of C, O, and N indicate that CRs with energies of a few GeV per nucleon --- which carry the bulk of the CR power --- pass through $\approx 10$ gm cm$^{-2}$ before escaping from the disk of the Galaxy. A mean disk density of one H atom cm$^{-3}$ gives $t_{esc} \approx 6\times 10^6$ yr, implying that $L_{CR} \approx 2\times 10^{40}$ ergs s$^{-1}$. Analysis of the composition of isotopic CR $^{10}$Be yields a larger value of $t_{esc}$, implying a smaller mean matter density but a larger effective trapping volume of the Galaxy so that, in either case,  
\begin{equation} 
L_{CR} \approx {u_{CR} V_{gal}\over t_{esc}} \approx 5\times 10^{40}\; {\rm ergs~s}^{-1}\;. 
\label{L_CR} 
\end{equation} 
 
The galactic SN luminosity $L_{SN} \approx (1$ SN$/30$ yrs$) \times 10^{51}$ ergs/SN $\approx 10^{42}$ ergs s$^{-1}$ which, even given a 10\% efficiency for converting the directed kinetic energy of SNe into CRs that seems feasible through the shock Fermi mechanism, is completely adequate to power the hadronic cosmic radiation. 
 
Although $\gamma$-ray astronomy was supposed to solve the cosmic-ray origin problem, this has not happened. The predicted $\pi^0$ decay feature at 70 MeV has not been detected from SNRs by EGRET \cite{esp96}, the Whipple imaging air Cherenkov telescope has not detected emission consistent with hadronic CR acceleration by SNRs \cite{buc98}, and the spectrum of the diffuse galactic $\gamma$-ray background contradicts the assumption that CR protons are uniformly distributed throughout the Galaxy with a spectral shape that is the same as observed locally \cite{hun97}. Moreover, it is becoming increasingly clear that the stochastic nature of explosive phenomena in the Galaxy is important for interpreting radiation emitted by Galactic CRs \cite{pe98,aws00}. 
 
The EGRET observations of the diffuse galactic $\gamma$ radiation voids the assumption that hadronic CRs uniformly inhabit the Galaxy, so that the required power could be overestimated if CR leptons emit most of the diffuse galactic radiation. Alternately, we could live in a region of enhanced CR hadron energy density compared to the Galactic average. This is more likely if CRs are produced by rare powerful events, which then the power requirements are also reduced. Our location in the Gould belt shows that we live near a region of enhanced stellar formation activity \cite{gp99}. 
 
Even if equation (\ref{L_CR}) is reliable, the massive stars that collapse to black holes while making GRBs can provide a kinetic power into the Galaxy comparable to that from SNe, as we now demonstrate. 
 
%*************************************************************
 
\section{Statistics of Cosmological GRBs} 
 
The new treatment of GRB statistics presented here is motivated by the recent Hipparchos, SN Ia and BOOMERANG results which indicate that we are living in a universe with non-zero cosmological constant \cite{pri00}, and by Beppo-SAX and follow-on observations demonstrating that GRBs are associated with events occurring in star-forming regions. For background on cosmological statistics, see \cite{wei72,der92,tot99,sch99,lr00,bd00}. 
 
The rate $\dot N(>\phi_p)$ at which observers detect events above some threshold peak flux $\phi_p$ is given by 
\begin{equation} 
\dot N(>\phi_p) = {4\pi c\over H_0} \int _0^\infty dz \; d^2_{\rm cosmo}(z) \;\dot n_{com}(z) P_{tr}[\phi_p(z)]\; ,
\label{dotN} 
\end{equation}   
where $H_0 = 3.24\times 10^{-18}h$ s$^{-1}$, $\dot n_{com}(z)$ is the comoving rate density (cm$^{-3}$ s$^{-1}$) of explosions at redshift $z$, and $\phi_p(z)$ relates the observed peak $\nu F_\nu$ flux to the peak source luminosity in the waveband at which the detector is most sensitive.  
 
The cosmological distance $d_{cosmo} \equiv d_L(1+z)^{-3/2} [q_\Lambda^0(z)]^{-1/2}$ is defined in terms of the luminosity distance $d_L = c(1+z) H_0^{-1}\int_0^z dz^\prime [q_\Lambda^0(z^\prime)]^{-1}$, where $q_\Lambda^0(z) \equiv \sqrt{(1+\Omega_m z)(1+z)^2 - \Omega_\Lambda(2z+z^2)}$. Results presented here are for an $\Omega_m = 0.3, \Omega_\Lambda = 0.7$ cosmology with $h=0.65$. Figure 1 shows $d_L$ and $d_{\rm cosmo}$ for this cosmology. 
 
\begin{figure}[t] 
\vskip-2.4in  
\centerline{\epsfxsize=0.66\textwidth\epsfbox{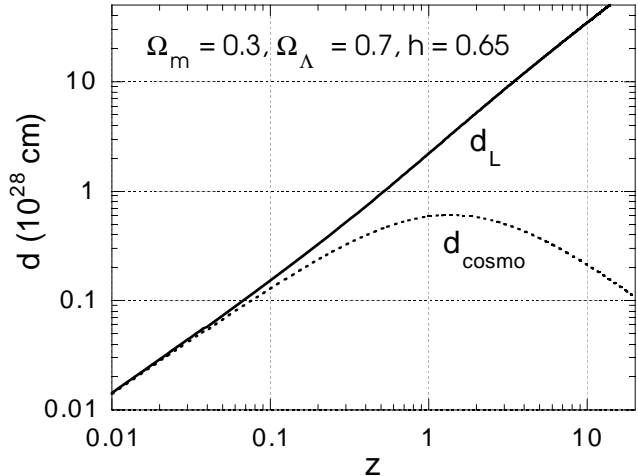}} 
\caption[] {Luminosity distance $d_L$ and cosmological distance $d_{\rm cosmo}$ in units of $10^{28}$ cm for a cosmology with $\Omega_m = 0.3, \Omega_\Lambda = 0.7$, and $h  = 0.65$.} 
\end{figure}  
 
The trigger efficiency $P_{tr}(\phi_{ph}) = \exp[-(\phi_{ph,0}/\phi_{ph})^{a_{ph}}]$ \cite{fis94,bd00}, where $\phi_{ph}$ is a photon flux. On the 1024 ms timescale, $\phi_{ph,0} \cong 0.26$ ph cm$^{-2}$ s$^{-1}$ and $a_{ph} = 5.3$. In the 50-300 keV band, the typical photon energy is $\sim 100$ keV when the peak of the $\nu F_\nu$ spectrum lies in the BATSE energy range. The $\nu F_\nu$ peak flux threshold is therefore $\phi_{p,0} \simeq 1.6\times 10^{-7}$ ergs $\times\phi_{ph,0} \simeq 4\times 10^{-8}$ ergs cm$^{-2}$ s$^{-1}$. Thus $P_{tr}[\phi_p(z)] \rightarrow H[\phi_p(z)-\phi_{p,0}]$, where H[u] is the Heaviside function such that $H(u) = 1$ for $u>0$ and $H(u) = 0$ otherwise. 
 
The flux density $S(\epsilon) = (1+z)d_L^{-2} \partial L[\epsilon(1+z)]/\partial \Omega$, where the directional spectral power at frequency $\nu = m_e c^2 \e/h$ is $\partial L(\epsilon)/\partial \Omega$. Although beaming can affect observable properties of a given GRB and the average event rate and total number of events in the Galaxy, it has no practical effect upon average power, because the product of the number of sources and energy or power per source remains unchanged. Only apparent isotropic energy releases and powers are quoted henceforth. 
 
The $\nu F_\nu$ flux threshold flux $\phi_p$ for a GRB with $S(\e )\propto \e^{-1}$ is given by the relationship $\phi_p = (4\pi d_L^2)^{-1} L_{peak} = (4\pi d_L^2)^{-1} L_0/\ln(\e_{max}/\e_{min}$). As we argue below, detectors are most sensitive to GRBs when the photon energy $\e_p^0$ of the peak of $\nu F_\nu$ flux is within the detector frequency window. In this approximation, the bolometric power $L_0$ exceeds $L_{peak}$ by the bandwidth correction factor $\ln(\e_{max}/\e_{min}$), which might be as large as $\sim 5$-10.  
 
\subsection{Clean and Dirty Fireballs} 
 
Consider a GRB detector that is most sensitive in a range centered about dimensionless photon energy $\bar\e$. The peak luminosity of a fireball is \cite{dcb99}  
\begin{equation} 
L_{peak} \simeq \eta E_0  
\, \cases{ t_d^{-1} ({\bar\e\over \e_p^0})^{-\delta}\;, & dirty~fireballs~with~$\bar\e \gg \e_p^0$ \cr\cr 
	 t_d^{-1} ({\bar\e\over \e_p^0})^{4/3}\;,& clean~fireballs~with~$\bar\e \ll \e_p^0$,\cr} 
\label{Lpeak} 
\end{equation} 
where $\e_p^0$ is the photon energy of the peak of the $\nu F_\nu$ spectrum during the prompt $\gamma$-ray luminous phase of a GRB, $E_0$ is the total energy released in the explosion, and $\eta$ is the efficiency to transform this energy into radiation during the deceleration timescale \cite{mr93} 
\begin{equation} 
t_d = {(1+z) x_d\over c\Gamma_0^2} = {1+z\over c\Gamma_0^2}\; ({3 E_0\over 4\pi \rho_0\Gamma_0^2})^{1/3} \cong 10(1+z) \;({E_{54}\over n_2})^{1/3} \Gamma_{300}^{-8/3}\;{\rm s}\;. 
\label{t_d} 
\end{equation} 
Equation (3) refers to a uniform circumburst medium with rest mass energy density $\rho_0  \cong  n m_pc^2$, where the proton density $n =100n_2$ cm$^{-3}$, $E_{54} = E_0/10^{54}$ ergs, and $x_d$ is the deceleration distance. The efficiency factor $\eta$ that relates $L_{peak}$ with $E_0$ and spectral parameters is \cite{dcb99} 
\begin{equation} 
\eta = {3(2g-3)\over 2g (3/4 + \delta^{-1})} \cong \cases{ 0.03 \;, & adiabatic~limit~with~$g=1.6$ \cr\cr 
	 0.25\;, & radiative~limit~with~ $g=2.9$,\cr}  
\label{eta} 
\end{equation} 
where $2+\delta$ is the photon number index of the prompt GRB spectrum at $\epsilon \gtrsim \e_p^0$, and the quoted efficiencies are for $\delta = 0.2$. The term $g$ defines the radiative regime according to the asymptotic behavior of the blast wave Lorentz factor $\Gamma(x) \propto (x/x_d)^{-g}$ when $x \gg x_d$. For a roughly adiabatic blast wave with $g=1.7$, as inferred from fits to the BATSE data \cite{bd00}, $\eta \sim 5$-10\%. 
 
In the external synchrotron-shock model \cite{tav96,coh97}, $\e_p^0 = q \rho_0^{1/2} \Gamma_0^4/(1+z)$, where $q$ is a parameter that takes into account the magnetic field energy density and efficiency to deposit swept-up energy into electrons. Assuming $q$ is constant with respect to changes in $\Gamma_0$, $E_0$, and $\rho_0$, then we can define the Lorentz factor $\bar\Gamma_0$ such that the observed peak $\nu F_\nu$ frequency is equal to $\bar \e$ . Hence $q\rho_0^{1/2}\bar\Gamma_0^4 = \bar \e$ defines $\bar \Gamma_0$, and we find that  
\begin{equation} 
L_{peak} \psim \eta E_0^{2/3}  
\, \cases{ \rho_0^{1/3+\delta/2} \Gamma_0^{8/3+4\delta}\;, & for~dirty~fireballs~with~$\Gamma_0 \lesssim \bar\Gamma_0$ \cr\cr 
	 \rho_0^{-1/3} \Gamma_0^{-8/3}\; ,& for~clean~fireballs~with~$\Gamma_0 \gtrsim \bar\Gamma_0$.\cr} 
\label{Lpeak1} 
\end{equation} 
 
As can be seen from equation (\ref{Lpeak1}), there are enormous selection biases against detecting dirty fireballs because the peak luminosity measured by the detector varies $\psim \Gamma_0^{3.5}$ when $\Gamma_0 \lesssim \bar \Gamma_0$. A steep reduction in the peak luminosity at photon energy $\bar\e$ likewise appears to be the case for clean fireballs, but the  shorter durations of the clean events (see eq.\ [\ref{t_d}]) mean that detectors trigger on fluence rather than peak flux when $\Gamma_0 \gg \bar \Gamma_0$. Thus GRB detectors are not so biased against the detection of clean as dirty fireballs, yet such detectors still preferentially detect those GRBs that produce prompt radiation with a $\nu F_\nu$ peak in the waveband at which the detector has greatest effective area. To avoid fine-tuning of $\Gamma_0$, one realizes that there must be two heretofore ``invisible" classes of explosive phenomena that remain to be discovered by detectors with appropriate design \cite{dcb99}. These considerations also resolve the apparent paradox that a beaming scenario should produce a wide range of $\e_p^0$, contrary to observations \cite{dbc99,bd00}. 
 
\subsection{Comoving Rate Density of Fireball Transients} 
 
Various lines of evidence lead to the conclusion that GRBs are formed in gaseous star-forming regions and are events related to the collapse of massive stars to black holes, possibly through the intermediate formation of a rapidly spinning neutron star (most similar to the supranova model \cite{vs98}). Calculations of GRB statistics are simplified when an assumption is made that the comoving FT rate density $\dot n_{com}(z)$ is proportional to the star formation rate (SFR) history of the universe \cite{tot97} as traced, for example, by Hubble Deep Field and SCUBA data.  
 
As the universe ages, we can expect star formation to be triggered by galactic mergers and interactions and therefore to be proportional to the density of galaxies. In this simple way of looking at things, the star formation rate is thus proportional to the proper galaxy density $n_{proper}\propto(1+z)^3 n_{com}$ near the present epoch.  At earlier times, the galaxy formation rate would tend to decline due to the time it takes to form the first structures, which depends on the primordial spectrum of density perturbations and the physics of galaxy formation. Star formation would therefore be impeded at recent times and at early times. 
 
This suggests that we parameterize the comoving rate density of FT sources by the expression 
\begin{equation} 
\dot n_{com}(z) = {\Sigma(1+a)\over [(1+z)^{-3} + a(1+z)^b]}\; , 
\label{ncom} 
\end{equation}  
which is normalized to $\Sigma$ at $z \rightarrow 0$. Figure 2 shows a range of parameterized models for the model SFR rate and, by assumption, the rate density history of explosions of a certain type. A model with $a = 0.03$ and $b=1$  provides a reasonable facsimile of the SFR curve reported in Ref. \cite{mpd98}.

\begin{figure}[t] 
\vskip-2.4in  
\centerline{\epsfxsize=0.66\textwidth\epsfbox{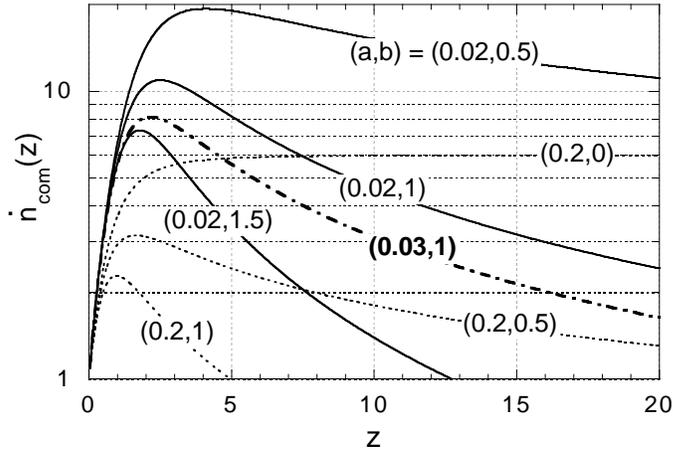}} 
\caption[] {Parameterization for comoving rate density of bursting sources for different values of parameters $a$ and $b$, and with $\Sigma$ set equal to unity. The curve with $a = 0.03$ and $b = 1$ is similar to the star formation rate as traced by UV radiation in the Hubble Deep Field \cite{mpd98}. } 
\end{figure}  
 
\subsection{GRB Redshift and Size Distributions} 
 
The left panel in Fig.\ 3 shows the observed redshift distribution of GRBs compiled from Jochen Greiner's tabulations\footnote{www.aip.de/$\sim$jcg/grb.html}. The right panel shows a model redshift distribution that is obtained by differentiating equation (\ref{dotN}) with respect to $z$, giving $d\dot N/dz = 4\pi c d^2_{\rm cosmo}(z) \dot n_{com}(z)/H_0$. This model assumes that the rate density of GRBs follows the parameterization of equation (\ref{ncom}) with $a=0.03$ and $b= 1$, and that all events are detected. In the absence of data from an extremely sensitive GRB telescope, this model is not directly comparable with data. Nevertheless, it can be seen that both distributions peak near $z \sim 1$-2 and display a tail extending to high redshifts. The observations suggest an excess of GRBs with low redshifts in comparison with the model, and this feature would be noted even if GRB 980425, associated with the SBb host galaxy of SN 1998bw at $z = 0.0085$, were excluded. The redshift distribution of GRBs with $z \lesssim 0.5$ is posed as a crucial discriminant between models \cite{der00,bd00,sch99}, because a large population of low luminosity, low-z GRBs (i.e., $dN/dz \sim const$ for $0 \lesssim z \lesssim 1$) requires a source rate density $\sim 2$-3 orders of magnitude greater than would be the case if there were no low luminosity  or low kinetic-energy GRBs. Although the FT rate would be much larger if a population of weak GRBs were discovered, the additional energy involved would be only a small fraction of the total energy released by GRBs with $E_{54} \gtrsim 10^{-2}$ \cite{der00}.  
 
\begin{figure}[t] 
\vskip-5.5in  
\centerline{\epsfxsize=1.0\textwidth\epsfbox{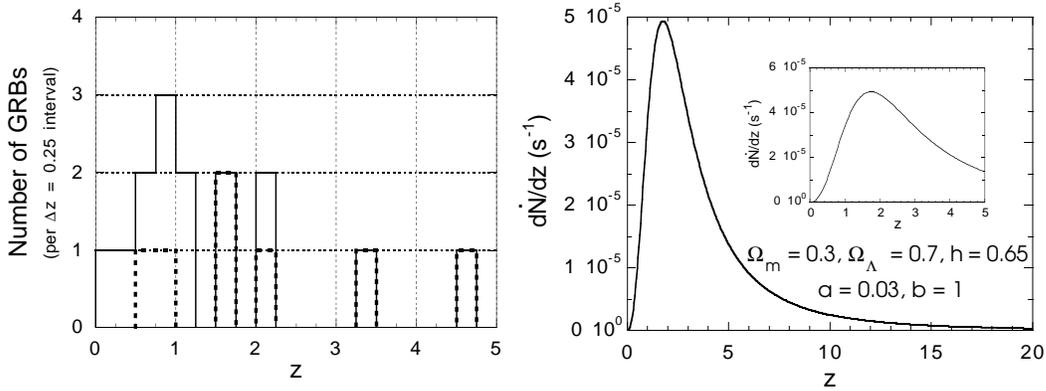}} 
\caption[] {(left) Distribution of GRBs with measured redshifts. Solid histogram represents all GRBs, and dotted histogram represents GRBs with $E_\gamma \gtrsim 10^{53}$ ergs. (right) Calculated distribution of redshifts assuming that the comoving event rate density is proportional to the star formation rate history of the universe as parameterized by equation (\ref{ncom}) with $\Sigma = 10^{-90}$ cm$^{-3}$ s$^{-1}$. Inset shows the low-$z$ distribution.} 
\end{figure}  
 
To understand this, we plot size distributions for different peak luminosities $L_{peak}$ in Fig.\ 4a using equations (\ref{dotN}) and (\ref{ncom}) with $a = 0.03$ and $b = 1$, using our chosen cosmology. These size distributions give the detection rate of GRBs that exceed peak $\nu F_\nu$ fluxes $\phi_p$, and are comparable to peak count-rate distributions reported by GRB detectors such as BATSE that trigger over a relatively narrow bandwidth when the count rate exceeds several $\sigma$ over background on some specified timescale. The bolometric peak luminosity $L_0 = {\cal B} L_{peak}$ \cite{der92} for a spectrum $\nu F_\nu\propto const$, where the bandwidth correction factor ${\cal B} = \ln(\e_{max}/\e_{min})\gtrsim 1$. Strong selection biases against detection come into play for sources that are detected at peak $\nu F_\nu$ frequencies $\e_p^0$ which are much different than the detector's sensitive frequency $\bar \e$ (Section IIIA). 
 
\begin{figure}[t] 
\vskip-5.2in  
\centerline{\epsfxsize=1.1\textwidth\epsfbox{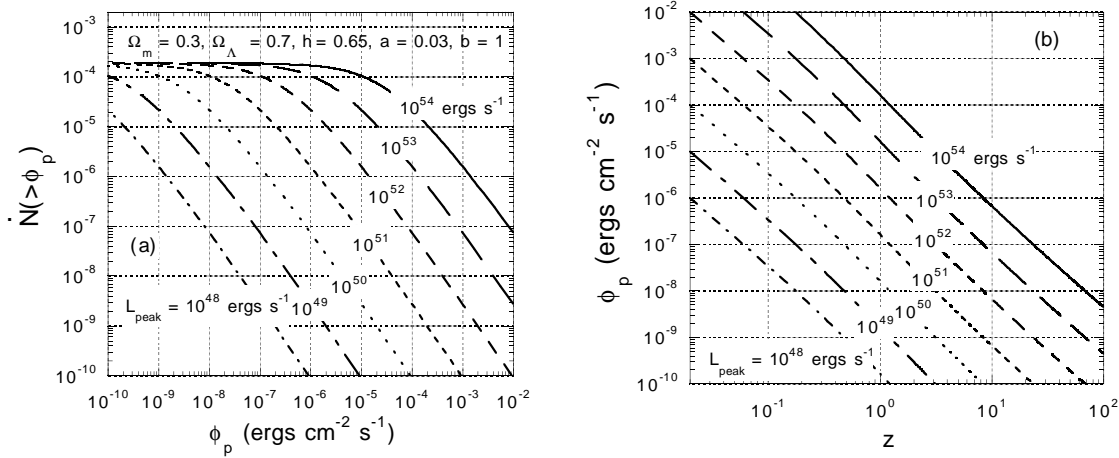}} 
\caption[] {(a) Size distribution of GRBs as a function of $\nu F_\nu$ peak flux $\phi_p$, for different peak luminosities $L_{peak}$ and with $\Sigma = 10^{-90}$ cm$^{-3}$ s$^{-1}$.  (b) Peak flux $\phi_p$ as a function of redshift $z$ for different values of $L_{peak}$.} 
\end{figure}  
 
The size distributions approach the Euclidean behavior $\dot N(>\phi_p)\propto \phi_p^{-3/2}$ at the bright end, and flatten to a constant at the dim end. The turnover flux in the size distribution occurs near the value of $\phi_p$ that corresponds to the peak $\nu F_\nu$ flux that would be detected from a source with brightness $L_{peak}$ at $z \cong 1$.  Fig.\ 4b verifies this. 
 
The redshifts of GRBs made possible by Beppo-SAX and the IPN (Interplanetary Network) and GCN (GRB Coordinates Network) display a range of peak luminosities of GRBs spanning at least 3 orders of magnitude. The redshifts of these sources are distributed about  $0.5 \lesssim z_p \lesssim 2$, though with unclearly defined wings at both low and high redshifts (Fig.\ 3a). To arrange a distribution of mixed peak-flux values in nearly the same redshift range requires a distribution $dN/d L_{peak}$ of $L_{peak}$ values that vary $ \psim L_{peak}^{-s}$ with $s \approx 3/2$. The pronounced change in slope of the peak count rate size distribution measured with BATSE  near count rates of $\approx 8$ ph cm$^{-2}$ s$^{-1}$ \cite{mee96} or $\phi_p \approx 10^{-6}$ ergs cm$^{-2}$ s$^{-1}$ implies from Fig.\ 4a that the $L_{peak}$ distribution is flatter than $-3/2$ at $L_{peak}\lesssim 10^{52}$ ergs s$^{-1}$ and steeper than $-3/2$ at $L_{peak}\gtrsim 10^{53}$ ergs s$^{-1}$.  The distribution cannot however be much flatter than $-3/2$ at the faint end unless we dismiss the association of GRB 980425 with SN 1998bw, or at least relegate it to a separate class. But this would then compromise the arguments that equate reddened optical excesses observed in the late-time optical afterglows of GRB 970228 and GRB 980326 with SNe emissions \cite{blo99,rei99,gal00}.  
 
Thus we assume that the peak luminosity distribution of GRBs follows a $dN/d L_{peak} \propto L_{peak}^{-3/2}$ behavior over a wide range of $L_{peak}$. To the extent that $L_{peak}$ is related to $E_0$ through an efficiency factor $\eta$, we can also parameterize the GRB $E_0$ distribution by $dN/d E_0 \propto E_0^{-3/2}$. This behavior was originally derived from  the numerical model \cite{bd00} but lacks a fundamental understanding at present, though it clearly reflects an underlying convex function describing the distribution of GRB events in terms of apparent energy release. 
 
\subsection{Fireball Transient Rate in an L$^*$ Galaxy} 
 
From Figs. 1-3 we see that the integrand in the calculation of the peak flux distribution in equation (\ref{dotN}) is dominated by events near $z_p \sim 1$-2, at least for sufficiently sensitive GRB detectors that reach threshold fluxes $\phi_{p,thr} \ll 10^{-6}$ ergs cm$^{-2}$ s$^{-1}$. From equation (\ref{dotN}), 
\begin{equation} 
\dot N(>\phi_p)\cong {4\pi c\over H_0} \int _0^{z_{max}(\phi_p)} dz \; d^2_{\rm cosmo}(z) \;\dot n_{com}(z)\simeq {4\pi c\over H_0} \cdot \Delta z \cdot d^2_{\rm cosmo}(z_p) \cdot \dot n_{com}(z_p)\;, 
\label{dotNap} 
\end{equation} 
which applies when the sensitivity threshold of the telescope is sufficiently good to detect sources at $z \gg z_p$. This appears to be the case for explosions yielding $\gamma$-ray energies $E_\gamma \gtrsim 10^{53}$ ergs that represent about 1/2 of the GRBs with measured redshifts (see Fig.\ 3a). The mean redshift of this sample is  $\approx 1.5$, and $\Delta z \cong 1$-2. This allows us to derive an absolute normalization on $\dot n_{com}(z_p)$ and therefore the present day comoving rate density of FTs that produce events with $E_\gamma \gtrsim 10^{53}$ ergs. 
 
From Fig. 1, we see that $d_{cosmo}(z_p) \cong 0.5\times 10^{28}$ cm. Writing the observed burst rate in terms of the frequency $\nu_B$ of observed events per day, we find from equation (\ref{dotNap}) that  
\begin{equation} 
\dot n_{com} (z_p) \simeq 3.0\times 10^{-90} \;{\nu_B h_{75}\over \Delta z} \; {\rm cm}^{-3}~{\rm s}^{-1} \; . 
\label{dotncom} 
\end{equation} 
 
The FT rate in an L$^*$ galaxy depends on its effective comoving volume. The density of L$^*$ galaxies in the local universe is $\sim 3\times 10^6$ Gpc$^{-3}$ \cite{der00}, so that an effective volume of an L$^*$ galaxy is $V_{L^*}\sim 10^{76} V_{76}$ cm$^3$, with $V_{76}\sim 1$. This number has a large uncertainty and could depend upon redshift.  Thus the rate at which an $L^*$ galaxy produces FTs with $E_\gamma \gtrsim 10^{53}$ ergs is  
\begin{equation} 
\dot N_* (E_\gamma > 10^{53} \;{\rm ergs};z)\cong \dot n_{com} V_{L^*}  \cong 3\times 10^{-14} \;{\nu_B h_{75}{\cal F}\over \Delta z} {\rm s}^{-1} \simeq 0.9 {\nu_B h_{75}{\cal F}\over \Delta z} {\rm ~GEM} \; ,
\label{NFT} 
\end{equation} 
where GEM is Galactic events per million years \cite{wij98}. The factor  
\begin{equation} 
{\cal F}(z) = {\dot n_{com}(z)\over \dot n_{com}(z_p)}  
\label{F} 
\end{equation} 
adjusts the comoving star formation rate averaged over the redshift range extending to $\approx 2\times z_p$ to the SFR rate occurring at a particular epoch $z$. To determine the event rate at the present epoch in the Milky Way, supposing it to be a typical L$^*$ galaxy,
\begin{equation} 
{\cal F}_0\equiv  {\cal F}(0) =({3\over ab})^{-3/(3+b)}+ a ({3\over ab})^{b/(3+b)}{ \over b= 1,a = 0.03}\hskip-10pt> 0.126. 
\label{F1} 
\end{equation} 
Note that the peak redshift $z_p$ of function (\ref{ncom}) is $z_p = -1+(3/ab)^{1/(3+b)}$. 
 
Two selection biases must be added into the accounting of equation (\ref{NFT}) to complete the problem. One is to account for the dirty and clean fireballs that do not trigger GRB detectors. The dirty fireball contribution is probably the more numerous unseen population, and the GRB statistical analysis in Ref.\cite{bd00} shows that $dN /d\Gamma_0 \propto \Gamma_0^{-0.25}$, for $\Gamma_0 \lesssim 260$, with $dN /d\Gamma_0$ falling off more steeply at larger values of $\Gamma_0$.\footnote{The density of the uniform external medium was set equal to 100 cm$^{-3}$, but the spectral model \cite{dcb99} is degenerate in the quantity $n_0 \Gamma_0^4$.} A simple integration gives the detection fraction of detectable transients compared to all FTs with $\Gamma_0 > 2$. If GRB detectors lose sensitivity to GRBs with $\Gamma_0 \leq$ 100 and 200, then the detection fraction is 0.5 and 0.18, respectively. This implies a clean and dirty fireball bias factor $\varrho_{\Gamma_0} \sim 2$-5. 
 
The second is to account for the environmental diversity that affects detection. Although blast wave physics shows that the emission properties scale to first order as $\sim \rho_0\Gamma_0^8$ \cite{dcb99}, the received flux gets washed out by light travel-time effects at low external medium densities, and by photoelectric absorption, Compton scattering and pair production attenuation \cite{bot99,db00} at high external medium densities. The low density bias seems unlikely given evidence that GRBs form in galaxies that are undergoing active star formation, but the high density bias is quite probable given recent results from Beppo-SAX \cite{ama00} and Chandra \cite{pir00} that seem to require a highly metal-enriched medium from SN events that precede by some months to years a GRB in the cases of GRB 990705 and GRB 991216, respectively. This is rather difficult to quantify, but an environmental biasing factor might be in the range $1 \lesssim \varrho_{\rho_0} \lesssim 10$, implying a total biasing factor $\varrho = \varrho_{\Gamma_0}\varrho_{\rho_0}\sim 2$-30. 
 
In view of these considerations, we find that events with $E_\gamma > 10^{53}$ ergs occur in a local L$^*$ galaxy within $z \lesssim 0.1$ at the rate 
\begin{equation} 
\dot N_*(E_\gamma > 10^{53}{\rm ergs})\; ({\rm GEM})\;  \cong  \;{\nu_B h_{75} V_{76} {\cal F}_0 \varrho\over \Delta z}\;\equiv 0.1\psi\;. 
\label{NGEM} 
\end{equation} 
Our best guess is that $\psi \sim 1$.  Equation (\ref{NGEM}) sets the normalization for the event rate of FTs in an L$^*$ galaxy such as the Milky Way. 
 
\subsection{Power and Number of FTs in the Milky Way} 
 
GRBs detected with energy $E_\gamma$ are produced by explosions with total energy $E_0\cong E_\gamma/\eta\simeq 10$-100 $E_\gamma$, using equation (\ref{eta}). We have argued in section IIIC that the $E_0$-distribution of GRBs can be approximated by 
\begin{equation} 
\dot N_*(E_0) = {K \over E_{54}} ({E_0\over E_{54}})^{-3/2} 
\label{dotNstar} 
\end{equation} 
where the exponent $-3/2$ is chosen in view of the observations which indicate a wide range of values of $E_\gamma$ at $z \lesssim 1$-2. We now adopt a notation where $\E$ refers to energy in units of $10^{54}$ ergs. Normalizing equation (\ref{dotNstar}) to the L$^*$ rate of GRBs and FTs with $\E_0 < \E < \E_1$ implies that 
\begin{equation} 
K = K(\eta,\E_0,\E_1) =  { \dot N_*(>\E_0)\over 2[\E_0^{-1/2} - \E_1^{-1/2}]}\;. 
\label{K} 
\end{equation} 
It now only remains to assign $\E_1$ and relate $\E_0$ to $\E_\gamma$ for a given efficiency $\eta$, which in turn relates $\dot N_*(E_0)$ to $\dot N_*(\E_\gamma > 0.1)$ from equation (\ref{NGEM}). 
 
The efficiency $\eta$ to convert the injected kinetic energy $E_0$ into $\gamma$-ray energy $E_\gamma$, given by equation (\ref{eta}), is assumed to be constant from event to event. We consider three efficiencies, namely $\eta$ = 100\%, 10\%, and 1\%,  which imply a maximum value of $\E_1$ given the beaming factor. The maximum energy available from the collapse of a neutron star, even a rapidly rotating 3 M$_\odot$ neutron star, is $\sim 10^{54}$ ergs. A beaming factor $\delta \Omega/4\pi \lesssim 1$\% seems ruled out by the data for the onset of the beaming break in some GRBs such as GRB 990510 and GRB 990123, as well as by the interpretation of fluorescence Fe Ly$\alpha$ emission \cite{pir00}. Thus we assume that the maximum apparent energy release is $\E_1 = 100$, corresponding to a 1\% beaming factor, and $\E_2$ = 10, corresponding to a 10\% beaming factor.  
 
Our grid of model runs therefore contains Runs 100a, 100b, 10a, 10b, and 1b, where the number in the label refers to the percentage value of $\eta$, and the letters a and b refer to $\E_1 = 10$ and 100, respectively. $\E_0$ takes values of 0.1, 1, and 10 for $\eta = 100, 10$, and 1\%, respectively. There is no Run 1a because $\E_0 = 10 = \E_1$.  
 
From $K$, we obtain the total power of FTs into the Galaxy from the expression 
\begin{equation} 
L_* \cong L_*(>\E_0) = K E_{54}\int _{\E_0}^{\E_1} d\E \cdot \E^{-1/2} \simeq 2\times 10^{54} K \E_1^{1/2}\;{\rm ergs~s}^{-1}\; 
\label{L*} 
\end{equation} 
for $\E_0 \ll \E_1$, and we obtain the total number of FT events from 
\begin{equation} 
N_*(>\E) \cong t_{gal} K\int _\E^{\E_1} d\E^\prime\;\E^{\prime -3/2} \simeq 2 t_{gal} K \E^{-1/2}\; 
\label{N*} 
\end{equation} 
for $\E \ll \E_1$, where the age of the Galaxy is given by $t_{gal} = 10^{10}t_{10}$ yr. 
 
Table 1 gives the results of $K$, the FT power $L_*$, and the total number of FT events $N_*$ during the life of the Milky Way for this grid of models. The value of $N_*$ is sensitive to $\E$ which ranges from $10^{-6} \lesssim \E \lesssim 10^{-3}$, corresponding to events with $10^{48} \lesssim E_0 $(ergs) $\lesssim 10^{51}$. The lowest value of $\E$ seems required if GRB 980425 is believed to be associated with SN 1998bw, implying a value $\E \lesssim 10^{-3}$. 
\begin{table} 
\caption{FT Power and Number of FT Events in an L$^*$ Galaxy such as the Milky Way} 
\label{table1} 
%\begin{tabular}{lrrr} 
%\begin{tabular}{lccc} 
\begin{tabular}{lddddd} 
   Run~~~ $\eta$& ~~$\E_0$~~ &  
   \multicolumn{1}{c}{ ~~~~~~~$\E_1$~~} ~~ & 
\multicolumn{1}{c}{~~~~~~~~~~$K/\psi$ (s$^{-1}$)} ~~& \multicolumn{1}{c}{~~~~~~~~~~($L_*/\psi$)\tablenote{Units of {\rm ergs~s}$^{-1}$}}~~& 
  \multicolumn{1}{c}{[$N_*/(t_{10}\psi)$]\tablenote{Rate for unbeamed outflows}}\\ 
\tableline 
100a~~100\% & 0.1 & 10 & 5.3$\times 10^{-16}$ & 3.3$\times 10^{39}$& $10^4\rightarrow 3\times 10^5$\\ 
100b~~~ $''$ & 0.1 & 100 & 4.9$\times 10^{-16}$ &  1.$\times 10^{40}$ & $ ''$ \\ 
10a~~~ 10\% & 1 & 10 & 2.1$\times 10^{-15}$ &  9.$\times 10^{39}$ & $3\times10^4\rightarrow 10^6$\\ 
10b~~~~ $''$ & 1 & 100 & 1.7$\times 10^{-15}$ &  3.$\times 10^{40}$ & $''$ \\ 
1b~~~~ 1\% & 10 & 100 & 6.9$\times 10^{-15}$ &  9.5$\times 10^{40}$& $10^5\rightarrow 3\times 10^6$\\ 
\end{tabular} 
\end{table} 
 
As can be seen from Table 1, FTs inject a time- and volume-averaged kinetic power between $\sim 10^{39}$-$10^{41}$ ergs s$^{-1}$ into the ISM, depending sensitively on the assumed efficiency for converting the directed kinetic energy of the outflow into soft $\gamma$ radiation. Comparing with equation (\ref{L_CR}), we see that FTs inject adequate power to produce the hadronic cosmic rays if the CR hadrons are accelerated by GRB blast waves with high efficiency, yet radiate gamma-rays with modest ($\eta \sim$1-10\%) efficiency. A value of $\eta \sim 5$\% is implied by calculations from the external shock model \cite{bd00}. A value of $\eta \sim 1$\%, as seems appropriate for an internal shock model, would require more power from GRB progenitors. However, no good statistical treatment has been performed within a wind scenario that can be used to assess detector biases, and these could change the value of $\psi$. For example, the beaming paradox resolved in Section IIIA in the external shock model remains unsolved to date by an internal shock model. But in either case, the progenitor stars of GRBs are seen to inject $\gtrsim 10^{40}$ ergs s$^{-1}$ into the ISM of an L$^*$ galaxy like the Milky Way.

\section{Cosmic Rays, Supernovae, and Fireball Transients} 
  
The most crucial test for a theory of cosmic-ray origin is to demonstrate that the putative source class provides the available power. The much rarer FTs and GRBs can provide this power because they are also much more energetic. An important point is that SNe are thought never to liberate more than a few $\times 10^{51}$ ergs of ejecta kinetic energy per explosion, with the bulk of the power that is potentially available in core-collapse SNe carried away by neutrinos. The massive stellar collapse events that make FTs evidently liberate as much as $10^{54}$ ergs per explosion in the supranova model; perhaps an order-of-magnitude more energy is possible in a collapsar model \cite{mfw99} or in the collapse of a massive Fe core to a black hole \cite{der00}. More likely, there is no strict uniformity of the properties of stellar progenitors that collapse to form black holes.  
 
\subsection{Ultra-High Energy Cosmic Rays} 
 
The energy density of UHECRs with particle energy $\gtrsim 10^{20}$ eV (see Fig.\ 9 in Ref.\ \cite{der00}) is
\begin{equation} 
u_U \sim \;{\rm few}\times 10^{-21} \;{\rm ergs~ cm}^{-3} \sim { L_* \over V_*}\tau_{p\gamma}\; 
\label{uhecr} 
\end{equation} 
where $\tau_{p\gamma} = \lambda_{p\gamma}/c$ is the timescale to degrade a particle's energy by a factor of $\sim 2$ through photopion production of UHECRs with the cosmic microwave background radiation. The term $L_*/V_*$ is the average power injected into an L$^*$ glaxy divided by the effective volume of an L$^*$ galaxy. For $10^{20}$ eV protons, the mean energy loss length is $\sim 140$ Mpc \cite{sta00}. Letting $L_* = 10^{40}L_{40}$ ergs s$^{-1}$, and recalling that the effective L$^*$ galaxy volume $V_{gal} = 10^{76} V_{76}$ cm$^3$, we obtain   
\begin{equation} 
{ L_* \over V_*}\tau_{p\gamma}\;\simeq 1.4\times 10^{-20}\;{L_{20} \over V_{76}}\;{\rm ergs~cm}^{-3}\;.
\label{L*overV*} 
\end{equation} 
If FTs and GRBs power the UHECRs, some modest but nonnegligible factor $\gtrsim 0.01$ of the available FT and GRB power must be processed into UHECRs to account for the observed energetics of metagalactic UHECRs. Waxman \cite{wax95} and Vietri \cite{vie95} first pointed out this coincidence.   
 
\subsection{Cosmic Rays between the Knee and the Ankle} 
 
The energy density of cosmic rays above the knee at $\cong 2$-4 PeV is  
\begin{equation} 
u_{\rm knee} \sim 10^{-16} {\rm~ergs~cm}^{-3}\;. 
\label{uknee} 
\end{equation} 
If GRBs and FTs power the PeV - EeV CRs, then CRs above the knee of the spectrum diffuse from the disk into the halo and then escape into intergalactic space.   
 
Rather than attempt to derive the halo escape timescale $\tau_H$ from a model for the Galactic halo, we instead invert the relation  
\begin{equation} 
{ L_* \over V_H}\tau_H\;\simeq 10^{-16}\;{\rm~ergs~cm}^{-3}\; . 
\label{lvh} 
\end{equation} 
to solve for $\tau_H$. Let the halo volume  $V_H = \pi r^2\ell$ and the escape timescale $\tau_H \cong k_H \ell/c$, where $r \cong 15$ kpc is the galaxy radius and $\ell$ is the effective halo height. Setting $u_{\rm knee} \simeq L_* k_H/(\pi r^2 c) = 5\times 10^{-17} L_{40} k_H$ ergs cm$^{-3}$, we obtain
\begin{equation} 
L_{40} k_H \simeq 10\;. 
\label{uc} 
\end{equation} 
This relation implies interesting properties concerning the halo diffusivity. If $L_{40}  \sim 1$ for the power into the PeV-EeV component, then the escape time of CRs near the knee of the spectrum is only an order-of-magnitude greater than the transit timescale, implying a  weak halo magnetic field.  
 
The high-mass progenitor stars of FTs could therefore power the PeV-EeV CRs if $\gtrsim $ 10\% of the energy of FTs is transformed into particles with these energies. The bend at the knee of the CR spectrum is due, in this interpretation, to galactic propagation effects.  
 
\subsection{GeV - PeV Cosmic Rays} 
The galactic cosmic rays between $10^9$-$10^{15}$ eV require, if distributed uniformly throughout the disk of the Galaxy, a power $\sim 5\times 10^{40}$ ergs s$^{-1}$ (Section II). For efficiencies $\sim 1$-10\% to transform the explosion energy into observed $\gamma$ rays, Table 1 shows that FTs and GRBs in our Galaxy could satisfy this power requirement, though the efficiency to transform the directed kinetic energy into CR energy must be high. Stars that collapse to black holes could therefore be the sources of the hadronic cosmic rays. 
 
The argument that the collapse of GRB and FT progenitor stars to black holes accelerates both the UHECRs and the GeV-EeV hadronic CRs is primarily based on simplicity --- this solution requires the fewest number of theoretical assumptions. A SN origin for the GeV-PeV Galactic CRs leaves open the question of cosmic rays with energies above the knee of the CR spectrum; FTs as the sole source class closes it. Compositional and theoretical studies near and above the knee of the CR spectrum will provide much of the necessary data to answer it. On the basis of available power from these explosions, it is therefore proposed that the high mass stars that collapse to form GRBs and FTs are the sources of the hadronic CRs.  
 
Other arguments for this cosmic-ray origin hypothesis are given in my paper written in memory of Jan van Paradijs \cite{der00}.  
 
\section{The Unidentified EGRET Sources} 
 
Apparently independent of the GRB problem looms the question of the identity of the $\sim 15$ COS B, $\gtrsim 100$ EGRET, and soon-to-be $\sim 10^3 (?)$ GLAST unidentified sources. These objects comprises an $\sim 3^\circ$ half-angle scale height disk component, an $\sim 20^\circ$ half-angle scale height mid-latitude population (which is wider and more extended than a bulge population), and an isotropic component representing $\lesssim 10$\% of the unidentified EGRET sources (UES). There is also detected a 10$^\circ$ diameter crown of a dozen or so EGRET $\gamma$ ray sources centered at $l=0^\circ,b=+5^\circ$, that is incidentally similar to a feature in COBE/DIRBE maps. 
 
Undoubtedly, some of the UES are pulsars, given our knowledge of Geminga. Grenier \cite{gre99} presents compelling evidence that some of the disk sources are coincident with X-ray plerions, which opens the possibility that several other low-latitude sources are members of this class. There may be separate populations of disk and mid-latitude sources \cite{geh00}, with the remaining isotropic fraction being some unidentified blazars of an extreme type, for example, COMPTEL MeV blazars or extreme blazars that put the $\nu F_\nu$ peak of the nonthermal synchrotron emission in the soft $\gamma$-ray band \cite{ghi99}. Some of the isotropic component could consist of the synchrotron self-Compton emissions of dirty fireballs \cite{dcm99} caught once in the act. Here the better imaging of GLAST will be enormously helpful. Each EGRET unidentified source is, of course, potentially a TeV source. 
 
On the basis of the spatial, spectral, and variability data, I \cite{der97} concluded that the solution to the UES which entails the least number of unproven assumptions is that these sources are isolated black holes accreting from the ISM. Only one population of such sources is required; the different scale-heights of the low-latitude and mid-latitude components represent younger ($\sim 10^7$-$10^8$ yr old) black holes found in dense molecular cloud complexes, and older black holes with large scale heights that are located within several hundred pc of the Solar system, respectively. 
 
The spectral and variability data argue against a large number of pulsars or isolated neutron stars comprising the UES if their emissions are like Geminga. The UES have soft spectra like blazars (photon index $\alpha \sim 2.3\pm 0.4$) rather than hard spectra like pulsars ($ \alpha \sim 1.7$, extending from 1.2-2.2). This argument cannot be pushed too far, though, as an underlying microquasar population has not been detected, and the spectral behavior of an isolated accreting black hole could be completely different from that of  a blazar or pulsar. Bondi-Hoyle accretion from the ISM onto $\sim 30$-100 $M_\odot$ black holes gives the correct order-of-magnitude luminosity for the $\sim 10^{35}$ ergs s$^{-1}$ disk component and the $\sim 10^{33}$ ergs s$^{-1}$  local component \cite{der97}. Ref. \cite{an99} points out that the luminosity relationship in advection-dominated models can alter the dependence on black-hole masses and Eddington rates, and advances this interpretation of the UES further. Punsly \cite{pun98} and G. Romero, at the Alicante INTEGRAL workshop, have a different approach based on magnetized black holes. 
 
A model based on isolated black holes accreting from the ISM requires a few $\times 10^{5}$-$10^6$ black holes with masses between $\sim 10$ and 100 $M_\odot$. Although this point requires more study, to satisfy compactness, variability and energy constraints, it is nearly certain that FTs with $\E_0 \gg 10^{-1}$ entail black hole formation. Whether the FTs with $\E_0 \ll 10^{-1}$ form black holes is not clear, but if they do, then as many as $10^6$-$10^7$ black holes are scattered throughout L$^*$ galaxies in an irregular distribution reflecting their birth and later movements. 
 
Table 1 demonstrates that the requisite number of black holes can be formed. A population of $\gtrsim 10$ M$_\odot$ black holes is formed in collapsar-type models \cite{bb94,tim96}. The black holes produced in the supranova model \cite{vs98} would have a few Solar masses unless delayed fallback of the ejecta onto the collapsing neutron star caused the black hole to become much more massive.  Moreover, a few Solar mass black hole found in the vicinity of its own ejecta from an earlier SN event could grow quite massive if the ejecta has not yet been fully dispersed. In a region with effective neutral hydrogen columns exceeding $\sim 10^{25}$ cm$^{-2}$ --- and assuming that the subsequent GRB did not completely disrupt the dense ejecta --- a nascent black hole could form an IR hotspot. Thus it may be possible even to have a population of moderate mass black holes in a supranova model. Whether this is consistent with IR source counts also requires further study.  
 
It hardly needs stated that AGILE, GLAST, and the new generation of TeV telescopes will make enormous progress in characterizing spectral, variability, and counterpart behavior of UES. Microlensing of background radiations to detect isolated black holes \cite{pac86} looks, upon initial inspection, extraordinarily difficult with present technology. Binary black hole systems such as Cygnus X-1 are well-known \cite{vpa95}; GRBs might announce their birth.

%**************************************************************
 
\section{The Lineage of Stars} 
 
The number of SN events that have occurred in the Milky Way is $\sim 10^{10}$ yr/$30$ yr $\approx 3\times 10^8$ and, as we have seen, FTs leave behind $\sim 10^4$ - $3\times 10^6$ black holes ranging in mass from a few Solar masses to tens of Solar masses. From what range of stellar masses do FTs originate?

\begin{table}[h] 
\caption{Rates of Different Nova Types in Supernova Units} 
\label{table2} 
\begin{tabular}{lddddd} 
   ~~& ~~$N1$~~ &  
   \multicolumn{1}{c}{ ~~~~~~~$N2$~~} ~~ & 
\multicolumn{1}{c}{~~~~~~~~~~$N3/N4$} ~~& \multicolumn{1}{c}{~~~~~~~~~$NF$\tablenote{Computed in \cite{der00}.}}~~& 
  \multicolumn{1}{c}{$~$}\\ 
\tableline 
Sbc-Sd & 0.21 & 0.86 & 0.14 & ~0.003& \\ 
\end{tabular} 
\end{table} 
 
This question can be answered by considering explosive event rates in Type Sbc-Sd galaxies,  using the tabulation of Ref.\ \cite{pan00}. (The Milky Way is Type Sbc, though with a bar of uncertain size.) The Supernova Unit is the number of events per $10^{10}$ Solar blue luminosities $L_{B,\odot}$ per century. If the Milky Way radiates 2$\times 10^{10}$ $L_{B,\odot}$, then Table 2 implies a time-averaged rate for all SNe types of 1 per 60 yr. Note the monotonic trend in Table 2, with the exception of the N1s (SN Ia's). The N1 rate is below the trend in the rates of core collapse novae because white dwarfs depend in most cases on being formed in binary systems to be driven into detonation. The ratio of the rates of N3/4s (SN Ib/Ic's) compared to the rate of all core collapse events (N2, N3, N4, and NF) is $\sim 0.14$. The ratio of the rates of NFs (GRBs and FTs) to all core-collapse events is $\sim 5.0\times 10^{-3}N_6$ for a $10^{10}$ yr lifetime of the Galaxy, assuming that FTs make $10^6 N_6$ black holes during this period. 
 
The differential stellar IMF (initial mass function) $\xi(\m)\propto \m^{-\sigma}$, where $\m$ is the mass of the zero age main sequence (ZAMS) star in units of Solar mass, and $\sigma$ is the differential IMF index. For a Salpeter IMF, $\sigma_{\rm S} = 2.35$, and $\sigma_{st} =2.8$ for a steep IMF. If $\m_{thr}$ is the mass threshold for initiating core collapse events, then the fraction of stars with mass $> \m$ is \cite{sw00} 
\begin{equation} 
\kappa \; = \; { \int_\m^\infty d\bar \m \;\xi (\bar \m) \over \int_{\m_{thr}}^\infty d\bar \m \;\xi (\bar \m)} \cong 0.14 {\rm ~and~} 5\times 10^{-3} N_6\;, 
\label{intmu} 
\end{equation} 
corresponding to the ratio of the rates of N3/N4s and NFs, respectively, to the rates of all core-collapse events. 
 
The relation $ \m = \m_{thr} \kappa^{1/{1-\sigma}}$ gives the lower limit to the ZAMS mass of stars that eventually collapse as different types of novae. The mass threshold for making SNe II/N2s is argued \cite{sw00} 
to be in the range  $\m_{thr} \cong 6$-14, depending in general upon metallicity.   
Stars with ZAMS masses  
\begin{equation} 
\m_{>N2/3} \cong  
 \cases{4.3 \m_{thr} ,&  $\sigma_{\rm S} = 2.35$ \cr 
        3.0 \m_{thr}\; , & $\sigma_{\rm st} = 2.8 $\cr}\;, 
\label{mN2/3} 
\end{equation}  
i.e., with masses $\gtrsim 20$-60 M$_\odot$, end their lives as N2s and N3s. According to this simple argument, stars with ZAMS masses  
\begin{equation} 
\m_{>NF} \cong  
 \cases{51 \m_{thr} N_6^{-0.74},&  $\sigma_{\rm S} = 2.35$ \cr 
        19 \m_{thr} N_6^{-0.56}\; , & $\sigma_{\rm st} = 2.8 $\cr}\;, 
\label{mF} 
\end{equation}  
i.e., stars with masses $\gtrsim 110$-300 M$_\odot$, end their lives as FTs. 
Similar results are obtained from the numerical calculations \cite{der00,bd00}. 
 
Based on these considerations, Fig.\ 5 sketches the scenario from the birth of stars on the stellar IMF with different ZAMS masses to their final fate as novae of various types, including the detonation supernovae, the NS core-collapse SNe, and the BH core-collapse hypernovae, namely GRBs and FTs. Observations and stellar structure calculations will test this picture, it being recognized that the actual situation is far more complex. See Ref.\ \cite{der00} for more discussion of this classification scheme and implications for different types of galaxies. 
 
\begin{figure}[b] 
\centerline{\epsfxsize=0.70\textwidth\epsfbox{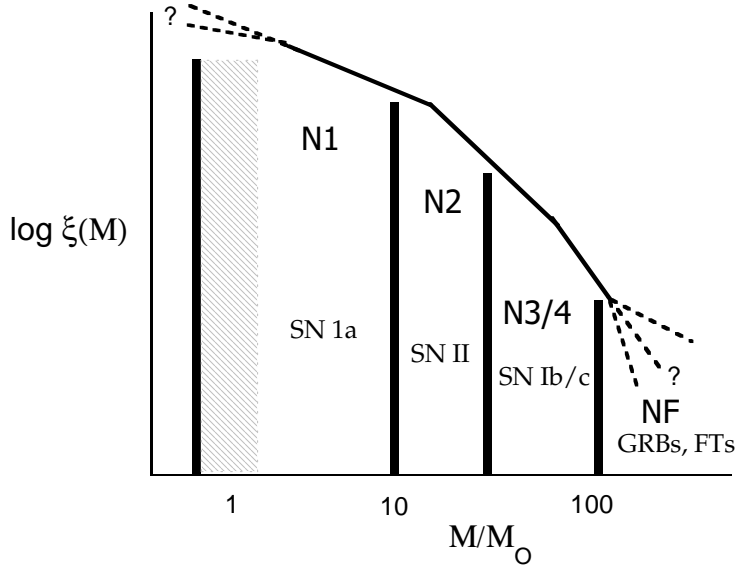}} 
\vskip-2.0in  
\caption[] {Sketch indicating the relationship between the stellar IMF $\xi(M)$, the ZAMS mass of the progenitor stars, and the explosion types that result from stars with different ZAMS masses. The shaded area represents the low ZAMS mass stars that evolve into white dwarfs on time scales exceeding the age of the Galaxy.} 
\end{figure}  
 
\section{Cosmic Rays from Gamma-Ray Bursts} 
 
The statistical estimates of available energy and power precede all subsequent considerations relating to a cosmic-ray source model. This contribution shows that the available power from the high mass stars that collapse to form GRBs and FTs is sufficient to power cosmic rays.  
 
Space limitations do not permit me to address the cosmic-ray acceleration and adiabatic loss problems here. Because they were such a strong focus of criticism of this CR origin hypothesis at the Heidelberg $\gamma$ 2000 workshop, however, it seems worthwhile to outline a proposed solution. 
 
Future writings \cite{dh00} will demonstrate that gyroresonant stochastic particle acceleration from turbulence in the relativistic blast waves of GRBs and FTs can accelerate protons and ions to ultra-high energies; that stochastic acceleration of particles by the evolving turbulence spectrum preferentially accelerates particles with a number index somewhat steeper than $\sim -2$; that $\gtrsim 10^{19}$ eV protons and ions can leak out of the blast wave during the prompt and afterglow phases to form the UHECRs, their Larmor radii being so large that they avoid adiabatic losses; that the bulk of the accelerated particles remain trapped in the blast wave throughout the prompt, afterglow, and nonrelativistic Sedov phase during which shock Fermi acceleration becomes more efficient than stochastic acceleration; that the late phases of a FT/GRB remnant represent in many respects a SNR except that higher energy particles are available from the acceleration that takes place as the relativistic blast wave decelerates; and that cosmic rays finally leave the remnant as it dissipates in the ISM, so that the adiabatic loss problem is solved (as in the standard model) by shock acceleration taking place during the expansion of the SNR until the energy density of the relativistic particle fluid is small compared to the magnetic field energy density of the ISM. 
 
Observations may well refute the cosmic-ray origin hypothesis put forward here. GLAST observations of a distinct $\pi^0$ feature in the vicinity of young SNRs could provide the $\gamma$-ray evidence that has been sought for decades. TeV telescopes could detect emission from SNRs that is more likely to have a hadronic than leptonic origin. On the other hand, GLAST or TeV observatories could detect hadronic emission at sites that arguably witnessed an earlier GRB, such as the Cygnus region, the Sco-Cen complex, or at sites in the direction of the $\sim 10^{18}$ eV cosmic ray anisotropies discovered by the AGASA \cite{tak99} and SUGAR arrays \cite{cla00}. 
 
In summary, the available evidence is not yet adequate either to verify or to reject a SN or GRB/FT origin of the hadronic cosmic rays.

\acknowledgments{I would like to thank H. V\"olk for criticisms of the acceleration model presented at the conference, and to acknowledge L. O'C. Drury, A. Levinson, A. Achterberg, and R. Lamb for useful and enjoyable discussions at the Heidelberg $\gamma$ 2000 workshop.  I also thank R. Schlickeiser, M. B\"ottcher, and R. Berrington for discussions and joint work. 
\vskip 0.3in 
}

\end{document}